\documentclass[aps,prd,showpacs,twocolumn,amsmath,10pt,superscriptaddress,floatfix,nofootinbib]{revtex4-1}
\usepackage{epsfig,amssymb,amsfonts,bm}
\usepackage{array}
\usepackage[active]{srcltx}
\usepackage{color}

\newcolumntype{L}{>$l<$} % math-mode version of "l" column type
\newcolumntype{C}{>$c<$} % math-mode version of "l" column type

\newcommand{\be}{\begin{equation}}
\newcommand{\ee}{\end{equation}}
\newcommand{\bea}{\begin{eqnarray}}
\newcommand{\eea}{\end{eqnarray}}
\newcommand{\beas}{\begin{eqnarray*}}
\newcommand{\eeas}{\end{eqnarray*}}

\newcommand{\al}{&}
\newcommand{\lag}{\mathcal{L}}
\newcommand{\A}{\mathcal{A}}

\newcommand{\itp}{\affiliation{CAS Key Laboratory of Theoretical Physics,
            Institute of Theoretical Physics,\\ Chinese Academy of Sciences,
            Beijing 100190, China}}
\newcommand{\bonn}{\affiliation{Helmholtz-Institut f\"ur Strahlen- und
             Kernphysik and Bethe Center for Theoretical Physics,\\
             Universit\"at Bonn,  D-53115 Bonn, Germany}}
\newcommand{\fzj}{\affiliation{Institute for
           Advanced Simulation, Institut f\"ur Kernphysik and
           J\"ulich Center for Hadron Physics,\\
           Forschungszentrum J\"ulich, D-52425 J\"ulich, Germany}}
\newcommand{\ific}{\affiliation{Instituto de F\'isica Corpuscular (IFIC),
             Centro Mixto CSIC-Universidad de Valencia,
             Institutos de Investigaci\'on de Paterna,
             Aptd. 22085, E-46071 Valencia, Spain}}
\newcommand{\ucas}{\affiliation{School of Physical Sciences,
            University of Chinese Academy of Sciences,
            Beijing 100049, China}}
\newcommand{\murcia}{\affiliation{Departamento de F\'\i sica,
           Universidad de Murcia, E-30071 Murcia, Spain}}

\begin{document}

\title{Towards a new  paradigm for heavy-light meson spectroscopy}

\author{Meng-Lin Du} \bonn
\author{Miguel~Albaladejo} \murcia
\author{Pedro Fernandez-Soler} \ific
\author{Feng-Kun~Guo}\email{fkguo@itp.ac.cn}\itp\ucas
\author{Christoph~Hanhart} \fzj
\author{Ulf-G.~Mei{\ss}ner} \bonn\fzj
\author{Juan~Nieves} \ific
\author{De-Liang~Yao}\ific

\begin{abstract}

Since 2003 many new hadrons, including the 
lowest-lying positive-parity charm-strange mesons
${D_{s0}^*(2317)}$ and ${D_{s1}(2460)}$,
were observed that do not conform with quark model expectations. 
It was recently demonstrated that various puzzles in the charm meson spectrum find a natural resolution, 
if the SU(3) multiplets for the lightest  scalar and axial-vector states, amongst them the 
${D_{s0}^*(2317)}$ and the ${D_{s1}(2460)}$, 
owe their existence to  the nonperturbative dynamics of 
Goldstone-Boson scattering off $D_{(s)}$ and $D^*_{(s)}$ mesons. 
Most importantly the ordering of the lightest strange and
nonstrange scalars becomes natural. 
In this work we demonstrate for the first time that this mechanism
is strongly supported by the
recent high quality data on the  ${B^-\to 
D^+\pi^-\pi^- }$ provided by the LHCb experiment.
This implies that the lowest quark-model positive-parity charm mesons, together 
with their bottom counterparts, if realized in nature, do not form the ground-state multiplet. 
This is similar to the pattern that has been established for the scalar mesons 
made from light up, down and strange quarks, where the lowest multiplet is considered to 
be made of states not described by the quark model. In a broader view, the hadron spectrum must be 
viewed as more than a  collection of quark model states.
\end{abstract}

% \pacs{xxxxx}

\maketitle

One of the currently most challenging problems in fundamental physics is to
understand the nonperturbative regime of the Quantum Chromodynamics (QCD), the
fundamental theory for the interaction of quarks and gluons.
However, since the quark and gluon fields are confined inside color-neutral hadrons,
what needs to be achieved is a quantitative understanding of
the hadron spectrum. First principle lattice QCD (LQCD) calculations are indispensable in this regard. In many cases, one can efficiently bridge the computationally intensive LQCD framework with complicated experimental observables using  chiral perturbation theory (ChPT)---the effective field theory for QCD at low energies---and its unitarization to fulfill probability conservation. 
In this work we demonstrate how such a combination leads to the resolution of a number of longstanding puzzles in charm-meson  spectroscopy. It also paves
the way towards a new paradigm in the spectroscopy for heavy-light mesons.

Until the beginning of the millennium heavy-hadron spectroscopy was assumed to
be well understood by means of the quark model~\cite{GellMann:1964nj,Godfrey:1985xj}, which
describes the positive-parity ground state charm mesons as bound systems of a
heavy quark and a light antiquark in a $P$-wave.
This belief was put into question in 2003, when the charm-strange scalar ($J^P=0^+$) and axial-vector ($1^+$) mesons
$D_{s0}^*(2317)$~\cite{Aubert:2003fg} and $D_{s1}(2460)$~\cite{Besson:2003cp}
were discovered (for recent reviews on new hadrons, see Refs.~\cite{Chen:2016qju,Chen:2016spr,Lebed:2016hpi,
Esposito:2016noz,Guo:2017jvc,Ali:2017jda,Olsen:2017bmm}), since 
the states showed properties at odds with the quark model.
Moreover, attempts to  adjust
the quark model raised more questions~\cite{Cahn:2003cw}. 
Various alternative proposals were put forward about the nature of these new
states including $D^{(*)}K$
hadronic molecules (loosely bound states of two colorless 
hadrons)~\cite{Barnes:2003dj,vanBeveren:2003kd}, 
tetraquarks (compact states made of two quarks and two antiquarks)~\cite{Chen:2004dy}
and chiral partners (doublets due to the chiral symmetry breaking of QCD in
heavy-light systems)~\cite{Bardeen:2003kt,Nowak:2003ra}. 
The situation became more obscure in
2004, when two new charm-nonstrange mesons, the $D_0^*(2400)$~\cite{Link:2003bd}
and $D_1(2430)$~\cite{Abe:2003zm}, were observed.
 In brief, the experimental discoveries brought up three puzzles:
\begin{enumerate}
  \item[(1)] Why are the $D_{s0}^*(2317)$ and $D_{s1}(2460)$ masses much lower than the quark model expectations for the lowest $0^+$ and $1^+$ $c\bar s$ mesons?
  \item[(2)] Why is the mass difference between the $D_{s1}(2460)$ and  the 
$D_{s0}^*(2317)$ equal to that between the ground state vector meson $D^{*+}$ and 
pseudoscalar meson $D^+$ within 2~MeV?
  \item[(3)] Why are the $D_0^*(2400)$ and $D_1(2430)$ masses almost equal to or even higher than their strange siblings, a relationship exploited in many works~\cite{Mehen:2005hc,Colangelo:2012xi,Alhakami:2016zqx,Cheng:2017oqh}, although states with an additional strange quark are typically 
  at least 100 MeV heavier since $m_s/m_d\simeq 20$, see, e.g., Ref.~\cite{Patrignani:2016xqp}?
\end{enumerate}
Although their bottom cousins are still being searched for in high-energy 
experiments, it is natural to ask whether such puzzles will be duplicated there 
and in other sectors.

{ As outlined below, in recent works it was demonstrated that analyses combining
effective field theory methods with LQCD allows one to resolve all those
puzzles. These analyses suggest that all low-lying positive-parity heavy open-flavor
mesons qualify as hadronic molecules. In this paper we add two crucial pieces to the existing line of reasoning,
namely we propose a lattice QCD study at unphysical quark masses that will allow one to see the two-meson character 
of the mentioned states more explicitly and we demonstrate that recent data on $B^-\to D^+\pi^-\pi^-$ show
a nontrivial structure fully in line with the proposed dynamical picture.}

One reason why the analyses that led to the $D_0^*(2400)$ and
$D_1(2430)$ resonance parameters in the Review of Particle Physics (RPP)~\cite{Patrignani:2016xqp} should be questioned is that the amplitudes used were inconsistent with constraints from the chiral symmetry of QCD. As its  chiral symmetry is 
 spontaneously broken, the pions, kaons and eta arise as Goldstone Bosons with derivative and thus energy-dependent interactions even for $S$-wave couplings. The standard Breit--Wigner (BW) resonance shapes used in the experimental analyses correspond, however, to constant couplings. Moreover, the energy range of these  states overlaps with various $S$-wave thresholds that necessarily need to be considered in a sound analysis, as these thresholds can leave a remarkable imprint on observables as will be shown below.
A theoretical framework satisfying such requirements is provided by the
unitarized chiral perturbation theory (UChPT) for heavy
mesons~\cite{Kolomeitsev:2003ac,Hofmann:2003je,
Guo:2006fu,Guo:2006rp,Gamermann:2006nm,Flynn:2007ki,Guo:2009ct,Liu:2012zya,Altenbuchinger:2013vwa,Du:2017ttu}. In this approach, ChPT at a given order is
used to calculate the interaction potential which is then resummed in a scattering
equation to fulfill exact two-body unitarity and allows for the generation 
of resonances as pioneered in Ref.~\cite{Truong:1988zp}.
Although there is no unique method for unitarization, different methods do not differ much as long as there are no 
prominent left-hand cuts~\cite{Pelaez:2015qba}, as is the case here.
It should be mentioned that any algebraic unitarization 
generates logarithms at higher order with wrong coefficients --- this is discussed for the case
of $\pi\pi$ scattering in Ref.~\cite{Gasser:1990bv} (for a more recent discussion see Ref.~\cite{Bruns:2017gix}).
As long as the unitarization is set up as for the amplitudes employed here, those appear only 
at orders higher than the order of the potential. However, there is no a priori way to estimate their
significance and the reliability of the amplitudes must be tested, e.g., by a comparison with lattice data
or experiment.
We will employ here the next-to-leading order (NLO) version
whose free parameters have been fixed to the Goldstone-Boson--charm-meson
scattering lengths determined in fully dynamical LQCD in channels without
disconnected diagrams~\cite{Liu:2012zya}. Later it was
demonstrated~\cite{Albaladejo:2016lbb} that these coupled-channel amplitudes
properly predict the energy levels generated in LQCD  (with a pion mass $M_\pi\simeq391$~MeV) for the isospin-1/2
channel even beyond the threshold~\cite{Moir:2016srx}.
This means that now the scattering amplitudes for the coupled 
Goldstone-Boson--charm-meson system are available that are based on 
QCD.
Moreover, those amplitudes allow us to identify the poles in the complex energy
plane reflecting the lowest positive-parity meson resonances of QCD in the charm sector as well
as in the bottom sector, once heavy quark flavor symmetry~\cite{Neubert:1993mb} is employed. 
The predicted masses for the lowest charm-strange positive-parity mesons are fully in line with the well-established measurements, and those for the bottom-strange mesons are consistent with LQCD results with an almost physical pion mass~\cite{Lang:2015hza}, see Table~\ref{tab:masses_cs} where the uncertainties quoted stem from the one-sigma
uncertainties of the parameters in the NLO UChPT determined in
Ref.~\cite{Liu:2012zya}.

It should be noted that in principle in addition to the mentioned uncertainty there is potentially
an additional uncertainty stemming from the truncation of the chiral expansion at NLO --- in other words terms of order $(M_K/\Lambda_\chi)^3\sim 10\%$ were neglected. It is
difficult to translate this uncertainty into an uncertainty of the pole locations since the parameters
of the NLO potential were fixed to LQCD data such that higher order effects are effectively 
absorbed into the parameters. However, we did check that a variation of the strength of 
the potential by 10\% simply moved the poles by 10-20 MeV --- but neither the number of poles
nor their sheets did change as a result of this variation.

\begin{table}[bt] \caption{Predicted masses of the lowest positive-parity heavy-strange mesons in comparison with the measured values~\cite{Patrignani:2016xqp} and latest LQCD results, in units of MeV.
}
\label{tab:masses_cs}
\begin{ruledtabular}
\begin{tabular}{C C C C}
  \text{~}    
  & \text{prediction} & \text{RPP} & \text{LQCD} \\ \hline
D_{s0}^* & 2315^{+18}_{-28}
& 2317.7\pm 0.6 & 2348^{+7}_{-4}~\text{\cite{Bali:2017pdv}} \\
D_{s1} &  2456^{+15}_{-21}  & 2459.5\pm0.6 &  2451\pm4~\text{\cite{Bali:2017pdv}}
\\
B_{s0}^*   & 5720^{+16}_{-23} & - & 5711\pm23~\text{\cite{Lang:2015hza}} \\
B_{s1} &  5772^{+15}_{-21}  & - & 5750\pm25~\text{\cite{Lang:2015hza}} \\ 
\end{tabular}
\end{ruledtabular}
\end{table}

\begin{table}[bt] \caption{Predicted poles corresponding to
the positive-parity heavy-light nonstrange mesons given
 as ($M,\Gamma/2$), with $M$ the mass and $\Gamma$ the 
total decay width, in units of MeV. The current
RPP~\cite{Patrignani:2016xqp} 
values are listed in the last column. 
}
\label{tab:poles}
\begin{ruledtabular}
\begin{tabular}{C C C C} 
  \text{~}    
  & \text{lower pole} & \text{higher pole} & \text{RPP} \\ \hline
D_0^* & \left(2105^{+6}_{-8}, 102^{+10}_{-11}\right)
& \left(2451^{+35}_{-26},134^{+7}_{-8}\right) & (2318\pm29,134\pm20) \\
D_1 &  \left(2247^{+5}_{-6}, 107^{+11}_{-10} \right)  & \left(2555^{+47}_{-30},
203^{+8}_{-9}\right) & (2427\pm40,192^{+65}_{-55})
\\
B_0^*   & \left(5535^{+9}_{-11},113^{+15}_{-17} \right) 
&  \left(5852^{+16}_{-19},36\pm5\right) & - \\
B_1 &   \left( 5584^{+9}_{-11}, 119^{+14}_{-17} \right) 
      &  \left(5912^{+15}_{-18}, 42^{+5}_{-4}\right) & -\\ 
\end{tabular}
\end{ruledtabular}
\end{table}

The first two of the puzzles listed above are solved in this picture: the  $D^{(*)}K$ hadronic molecules do not correspond to the quark model $c\bar s$ states; spin symmetry predicts that the binding energies  are independent of the heavy meson spin up to an uncertainty of about $10\%$, as the leading spin symmetry breaking interaction is also of NLO in the chiral expansion. Moreover, 
there are two poles, corresponding to two resonances, in the $I=1/2$ and strangeness $S=0$ channel. The predicted poles, located at the complex energies $M-i\, \Gamma/2$,
 for both scalar and axial-vector charm and bottom mesons are listed in
Table~\ref{tab:poles}. The masses for the lower nonstrange resonances are smaller than those for the strange ones, leading to the solution to the third puzzle. For comparison the currently quoted  
masses and widths of
the $D_0^*(2400)^0$ and $D_1(2430)^0$ given in RPP are also listed.

This pattern of two $I=1/2$ states emerges naturally in the underlying formalism  since already  leading order ChPT interactions are  attractive in two flavor multiplets, to which the two states belong: the anti-triplet and the sextet~\cite{Kolomeitsev:2003ac,Albaladejo:2016lbb}. These  two scalar isospin $I=1/2$ states were predicted in
the earlier works of Refs.~\cite{Kolomeitsev:2003ac,Guo:2006fu}, where, however, less refined amplitudes were employed.

\begin{figure}[tb]
  \centering
   \includegraphics[width=\linewidth]{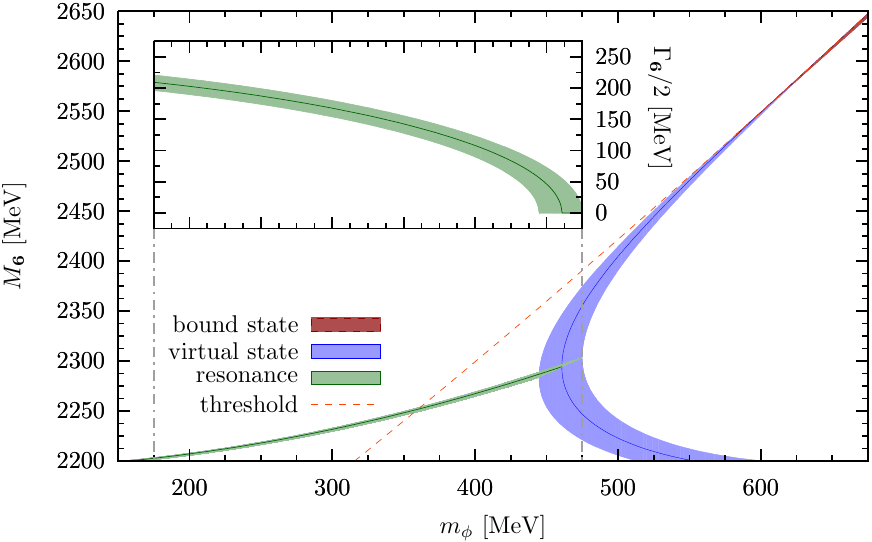}
   \caption{Illustration of the mass of the predicted sextet state at the SU(3) symmetric point as a function of the Goldstone boson mass $m_\phi$.  Below $m_\phi\lesssim475$~MeV, the pole is a resonance with its imaginary part ($\Gamma_{\bm{6}}/2$) shown in the inserted sub-figure. Above $m_\phi\simeq 475$~MeV, it  evolves into a pair of virtual states, and finally it becomes a bound state at $m_\phi\simeq 600$~MeV. }
  \label{fig:sextet_SU3}
\end{figure}

{ Given the above discussion, it is important to test the scenario outlined above as much as possible. In this work we discuss
two possible paths: On the one hand we propose a numerical experiment on the lattice, on the other hand we demonstrate that
recent experimental data provide additional support of the nontrivial dynamics that leads to the existence of the light
positive-parity open-charm states.

If the mentioned states were $\bar q c$ states, they would all be members of the flavor anti-triplet --- the presence of the sextet is a
nontrivial prediction emerging from the meson-meson dynamics that the picture presented above is based on.}
We notice that while we predicted two $I=1/2$ states (see Table~\ref{tab:poles}),  the Hadron Spectrum Collaboration reported only one, located just 
below the  $D\pi$ threshold, in their lattice calculation at $M_\pi\simeq391$~MeV~\cite{Moir:2016srx}. This is in line with the expectation that the lower pole 
would become a bound state at $M_\pi\gtrsim350$~MeV~\cite{Liu:2012zya}. { The authors of Ref.~\cite{Moir:2016srx} report that they ``do not find
any further poles in the region where ... [their lattice analysis] constrained the amplitudes''.  This does on the other hand not exclude the presence of the second
pole advocated in Ref.~\cite{Albaladejo:2016lbb} as well as above: The quote simply reflects the fact that while various of the amplitudes employed in  the
analysis of Ref.~\cite{Moir:2016srx} contained a second pole, its location was strongly parametrization-dependent~\cite{Davidprivate}. 
With the quark masses used in Ref.~\cite{Moir:2016srx}, the predicted sextet pole is located deep in the complex plane~\cite{Albaladejo:2016lbb}, and thus it is 
not captured easily.
The advantage of our
amplitudes compared to those employed in the analysis of Ref.~\cite{Moir:2016srx} is that they are constrained by both the pattern of chiral symmetry
breaking of QCD as well as lattice data in additional channels. To further test our explanation for the light positive parity open charm states,}
 we propose to search for them in lattice studies at a SU(3) symmetric point, with a relatively large quark mass $m_u=m_d=m_s$,  such that the 
 lightest pseudoscalar-meson masses will be near or above $m_\phi\gtrsim475$~MeV. We predict that the sextet pole will become a virtual state below 
 threshold for such large quark masses, and that it would even become a bound state for higher quark masses. This behaviour is illustrated 
  in  Fig.~\ref{fig:sextet_SU3}, where one can see that  now the pole is  close to  threshold, and  it should be easy to detect in a lattice calculation.
{ Note that the trajectory of the pole displayed in Fig.~\ref{fig:sextet_SU3}, in particular that in a certain parameter range resonance poles
exist in the complex energy plane below threshold,  is common for two-meson states in a relative $S$-wave. This feature is discussed in quite
general terms in Ref.~\cite{Hanhart:2014ssa} (see also Refs.~\cite{Hanhart:2008mx,Albaladejo:2012te} for the $f_0(500)$ case) and was first presented for the open flavor states in the focus here in Ref.~\cite{Guo:2009ct}.}

In the following, we show that our resolution to these puzzles is backed by precise experimental data by showing that the amplitudes with the two  $D_0^*$ states
are fully consistent with the LHCb measurements of the reaction $B^-\to
D^+\pi^-\pi^-$~\cite{Aaij:2016fma}, which are at present the best data providing
access to the $D\pi$ system and thus to the nonstrange scalar charm
mesons. Therefore, all the available theoretical,
experimental and LQCD knowledge is consistent with the existence of two $D_0^*$ states in the mass region where there was believed to be only one $D_0^*(2400)$.

\begin{figure}[tb]
  \centering
  \includegraphics[width=\linewidth]{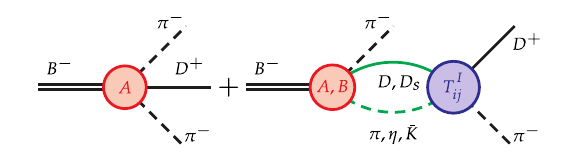}
 \caption{The decay amplitude for $B^-\to D^+\pi^-\pi^-$.
  Here, $A,B$ parameterize the production vertex, see
  Eq.~\eqref{eq:amp}, and
  $T_{ij}^{I}$ denotes the final state interactions between the charm and
  the light  mesons.
}
\label{fig:feyn}
\end{figure}

The Feynman diagrams for the decay amplitude for $B^-\to D^+\pi^-\pi^-$ are shown in
Fig.~\ref{fig:feyn}.
All the channels ($D^+\pi^-$, $D^0\pi^0$, $D^0\eta$ and $D_s^+ K^-$)
coupled to $D^+\pi^-$ need to be considered in the intermediate states.
The decay amplitude in the energy region up to 2.6~GeV, which is sufficient to study the
low-lying scalar states,
can be decomposed into $S$-, $P$- and $D$-waves,
\begin{eqnarray}
 \A(B^- \to D^+\pi^-\pi^-) = \sum_{L=0}^2
  \sqrt{2L+1} \A_L(s) P_L(z)  \, ,
\end{eqnarray}
where $\A_{0,1,2}(s)$ correspond to the amplitudes with $D^+\pi^-$ in the $S$ , $P$  and
$D$ waves, respectively, and  $P_L(z)$ are the Legendre polynomials. For the $P$- and $D$-wave amplitudes
we use the  same BW form as in the LHCb analysis~\cite{Aaij:2016fma}.
However, for the $S$-wave we employ
\begin{eqnarray}
  \A_{0}(s) &=& A \bigg\{ E_\pi \!\left[ 2+
  G_{1}(s)\left(\frac53 T^{1/2}_{11}(s) + \frac13
  T^{3/2}(s) \right) \right] \nonumber\\ 
  &&+ \frac13 E_{\eta} G_{2}(s) T^{1/2}_{21}(s)
  + \sqrt{\frac23 } E_{\bar K} G_{3}(s) T^{1/2}_{31}(s) \bigg\}
  \nonumber\\
  && + B E_\eta
  G_{2}(s) T^{1/2}_{21},
\label{eq:amp}
\end{eqnarray}
where 
$A$ and $B$ are two independent couplings following from SU(3) flavor symmetry~\cite{Savage:1989ub},
and $E_{\pi,\eta,\bar K}$ are the energies of the light mesons. { The effective Lagrangian for the production vertex leading to the above amplitude can be found in Appendix~\ref{app:Lagrangian}.} Here the
$T^{I}_{ij}(s)$ are the $S$-wave scattering amplitudes for the
coupled-channel system with total isospin $I$,
where $i,j$ are channel indices with $1,2$ and 3 referring to $D\pi$, $D\eta$ 
and $D_s\bar K$, respectively.
 These scattering amplitudes can be found in
Ref.~\cite{Liu:2012zya} where also all the parameters were fixed.
The unitarity relation
\begin{equation}
  \text{Im}\, \A_{0,i}(s) =  - \sum_{j} T^*_{ij}(s) \rho_j(s)^{} \A_{0,j}^{}(s),
  \label{eq:unitarity}
\end{equation}
with $\rho_j(s)$ the two-body phase space
factor in channel-$j$, is satisfied as long as $\text{Im}\, G_{i}(s) = -\rho_i(s)$,
which allows us to represent $G_{i}(s)$ via a once-subtracted dispersion relation~\cite{Oller:1998zr}. The 
same subtraction constant $a_A$ is taken for all channels.
The amplitude of Eq.~(\ref{eq:amp}) embodies chiral symmetry constraints and 
coupled-channel unitarity, and thus has a sound theoretical foundation. Here the final state interaction between the two $\pi^-$ mesons is neglected，
because the two pions are in an isospin-tensor state, and they have a large relative momentum so that they quickly fly away from each other.

\begin{figure*}[tbh]
  \begin{center}
   \includegraphics[width=0.31\linewidth]{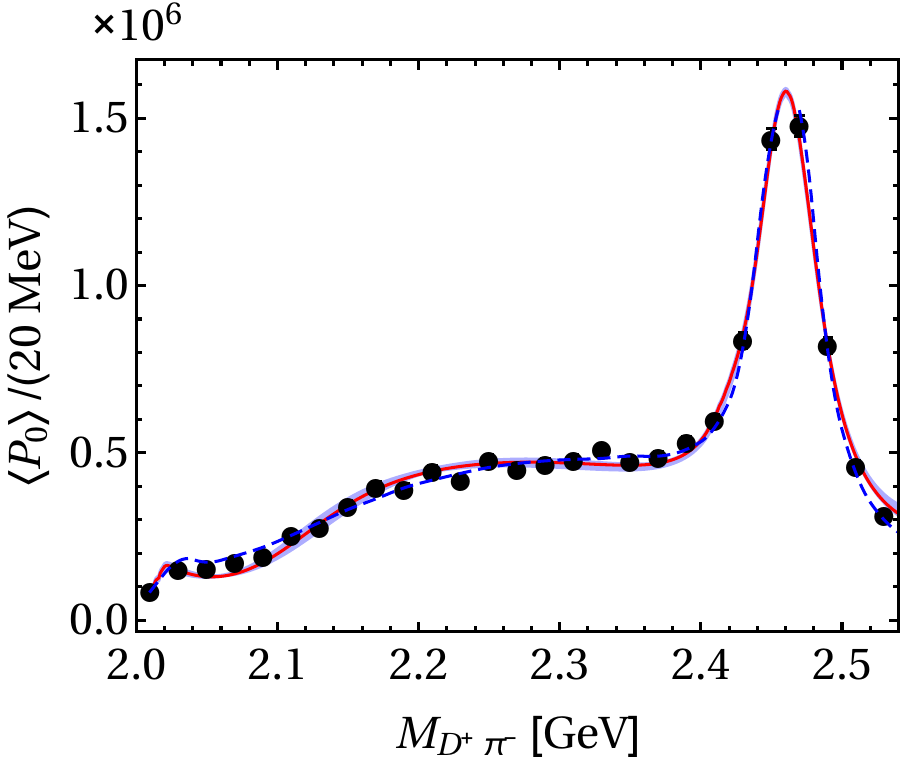} \hfill
  \includegraphics[width=0.32\linewidth]{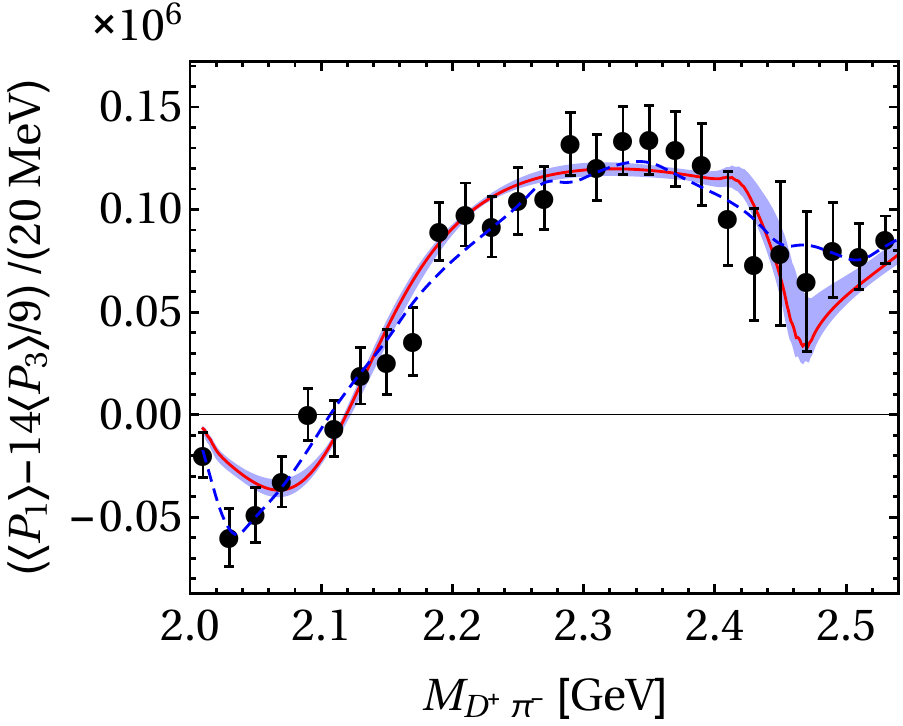} \hfill
  \includegraphics[width=0.31\linewidth]{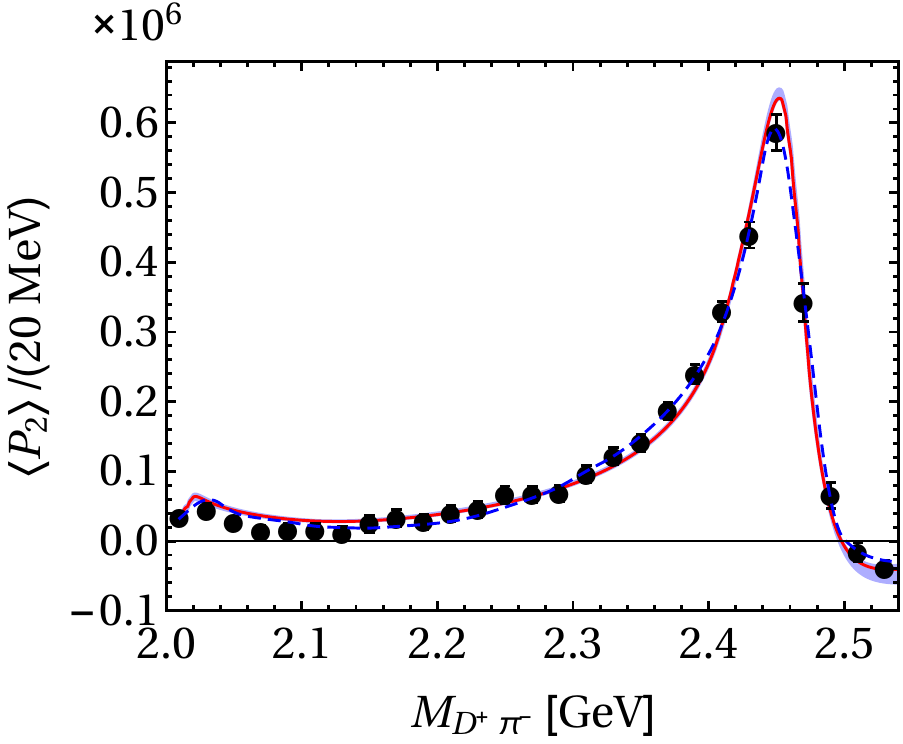}
 \end{center}
  \caption{
  Fit to the LHCb data for the angular moments $\langle P_0\rangle$,
  $\langle P_1\rangle - 14\langle P_3\rangle/9$ and $\langle P_2\rangle$ for the  $B^-\to
D^+\pi^-\pi^-$ reaction~\cite{Aaij:2016fma}. 
The largest error among $\langle P_1\rangle$ and $14\langle P_3\rangle/9$ in each bin is taken as
the error of $\langle P_1\rangle - 14\langle P_3\rangle/9$. The solid lines show  
our results, with error bands corresponding to the one-sigma uncertainties propagated 
from the input scattering amplitudes, while the
dashed lines stand for the LHCb fit using cubic splines for the $S$-wave~\cite{Aaij:2016fma}.
\label{fig:dpi}}
\end{figure*}

The so-called angular moments, see, e.g.~\cite{Aaij:2014baa,Aaij:2016fma}, contain important information about the partial-wave phase variations. 
Neglecting partial waves with $L\geq3$, which is perfectly
fine in the energy region of interest as indicated by the LHCb data, the first few moments are given by
\begin{eqnarray}
  \al\al\langle P_0\rangle \propto |\A_0|^2 + |\A_1|^2 + |\A_2|^2 \,,\nonumber\\
  \al\al\langle P_2\rangle \propto \frac25|\A_1|^2 + \frac27|\A_2|^2
+\frac{2}{\sqrt{5} } |\A_0||\A_2| \cos(\Delta\delta_2) \, ,\nonumber\\
\al\al  \langle P_{13}\rangle \equiv \langle P_1\rangle -\frac{14}{9} \langle
P_3\rangle \propto \frac2{\sqrt{3}} |\A_0||\A_1| \cos(\Delta\delta_1) \,,
\label{eq:moments}
\end{eqnarray}
where $\Delta\delta_{1,2}$ are the phase differences of $P$- and $D$-waves
relative to the $S$-wave, respectively. Instead of $\langle P_1\rangle$ and $\langle
P_3\rangle$ we propose to analyze the linear combination $\langle P_{13}\rangle $,
since it only depends on the $S$-$P$
interference up to $L=2$ and is particularly sensitive to the
$S$-wave phase motion.

We fit to the data  of the moments defined in Eq.~\eqref{eq:moments} up to $M_{D^+\pi^-}=2.54$~GeV 
for the decay  $B^-\to D^+\pi^-\pi^-$ measured by the LHCb
Collaboration~\cite{Aaij:2016fma}. Except for the $S$-wave $D\pi$ given in
Eq.~\eqref{eq:amp}, we include the resonances $D^*$ and $D^*(2680)$ in the $P$-wave and $D_2(2460)$ in the $D$-wave. Their masses and widths are fixed to the
central values in the LHCb analysis~\cite{Aaij:2016fma}, and their phase
parameters are denoted by $\delta_{D^*}$, $\delta_{D^*}'$ and $\delta_{D_2}$,
respectively. The best fit has $\chi^2/{\rm d.o.f.}=1.7$ and the parameter values are $B/A = -3.6\pm 0.1$,
$a_A = 1.0\pm 0.1$,  $\delta_{D^*} = -0.42\pm0.04$, $\delta_{D^*}' = 1.1\pm0.2$,
and $\delta_{D_2} = -0.83\pm0.07$.
We do not show the four normalization parameters
(three for these resonances and one for the $S$-wave).  A comparison of the best fit with 
the LHCb data is shown in Fig.~\ref{fig:dpi} together with the best fit provided by 
the LHCb collaboration~\cite{Aaij:2016fma} (dashed) { where cubic splines were used to interpolate
between certain anchor point --- below we detail this method further.} 
The bands in Fig.~\ref{fig:dpi}
reflect the one-sigma errors of the parameters in the scattering amplitudes  
determined in Ref.~\cite{Liu:2012zya}.
It is worthwhile to notice that in $\langle P_{13}\rangle$, where the
$D_2(2460)$ does not play any role, the data show a significant variation
between 2.4 and 2.5~GeV. Theoretically this feature can now be understood as a
signal for  the opening of the $D^0\eta$ and $D_s^+ K^-$ thresholds at 2.413 and 2.462~GeV, respectively,
which leads to two cusps in the amplitude. { This effect is amplified by the higher pole which is relatively
close to the $D_s\bar K$ threshold on the unphysical sheet.}

\begin{figure}[tbh]
  \begin{center}
   \includegraphics[width=0.9\linewidth]{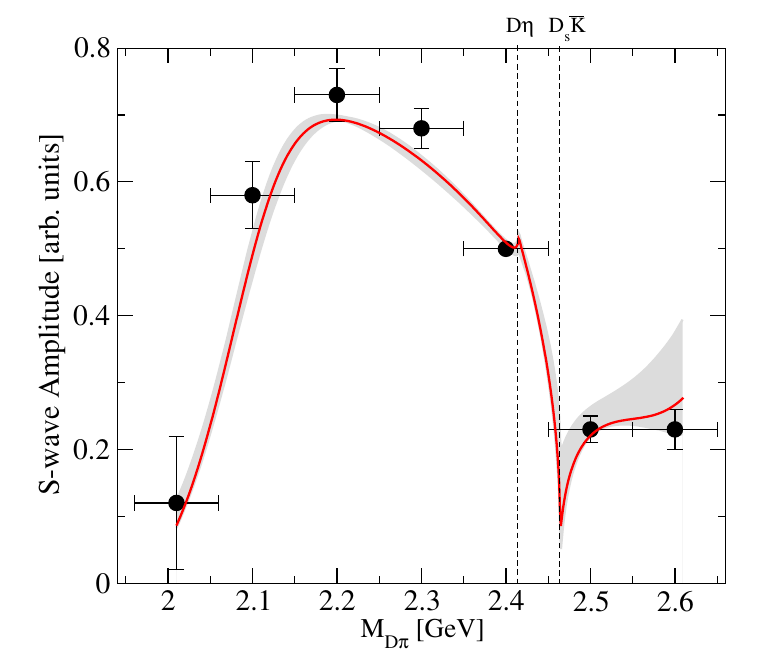} 
 \end{center}
  \caption{Comparison of the $S$-wave amplitude determined in this work to the $S$-wave anchor points found in the 
  experimental analysis, shown as the data points~\cite{Aaij:2016fma}. The red line gives the best fit results and the
  grey band quantifies the uncertainties that emerged from the fitting procedure. The fitting range extended up to
  2.55~GeV. The dashed perpendicular lines indicate the location of the $D\eta$ and $D_s \bar K$ threshold, respectively.
\label{fig:dpiSwave}}
\end{figure}

{ One might wonder if the discrepancy between our amplitude and the data
for $\langle P_{13}\rangle$ at low energies points at a deficit of the former. 
Fortunately the LHCb Collaboration provides more detailed information on their $S$-wave
amplitude in Ref.~\cite{Aaij:2016fma}: In the analysis of the data a series of anchor points
were defined where the strength and the phase of the $S$-wave amplitude were extracted
from the data. Then cubic splines were used to interpolate between these anchor points.
In  Fig.~\ref{fig:dpiSwave} the $S$-wave amplitude fixed as described above is compared to the
LHCb anchor points. Not only shows this figure very clearly that the strength of the $S$-wave amplitude
largely determined by the fits to lattice data is fully consistent with the one extracted from the data for $B^-\to D^+\pi^-\pi^-$,  
the  amplitude fixed in experiment also shows 
clear structures at both the $D\eta$ and $D_s \bar K$ thresholds. From our point of
view the most natural explanation of those structures
is that they are the mentioned cusps enhanced in impact by the pole located nearby. 
Thus the comparison of the
$S$-wave amplitude extracted by the LHCb Collaboration with our result
shows the role of the higher pole in the $I=1/2$ and $S=0$ channel even more clearly
than the angular moments discussed above. 
This clearly highlights  the importance of  a coupled-channel treatment for this
reaction. An updated analysis of the LHC Run-2 data is called for to confirm the prominence of the
two cusps.
Notice that the shape of the $S$-wave is determined by only two real parameters
($B/A$, $a_A$), while its phase motion is largely determined from
unitarity, Eq.~\eqref{eq:unitarity}.

\begin{figure*}[tbh]
  \begin{center}
  \includegraphics[width=0.31\linewidth]{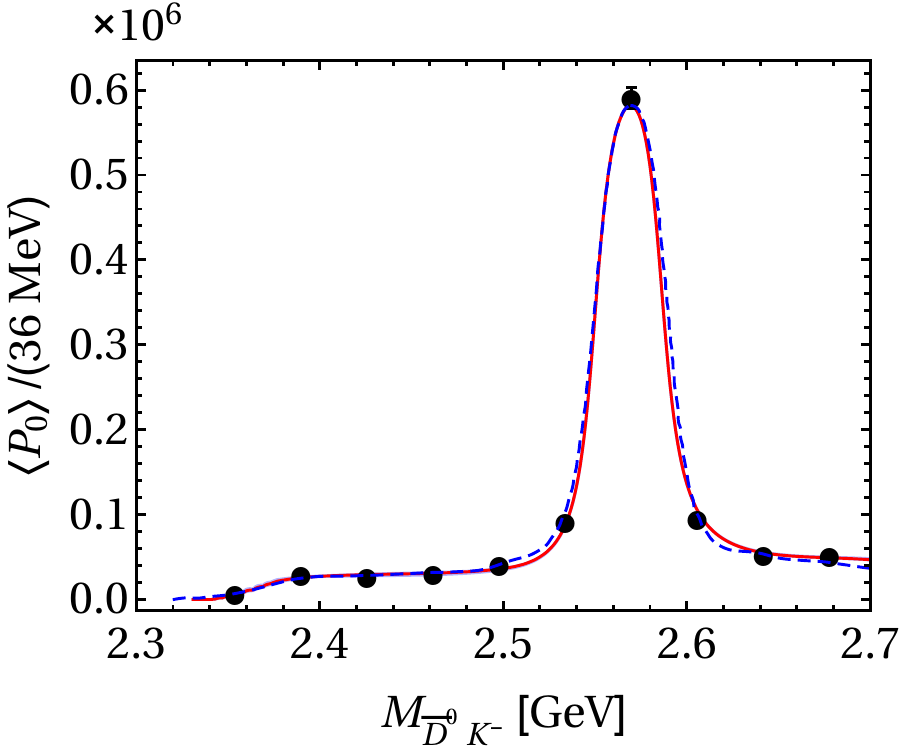} \hfill
  \includegraphics[width=0.32\linewidth]{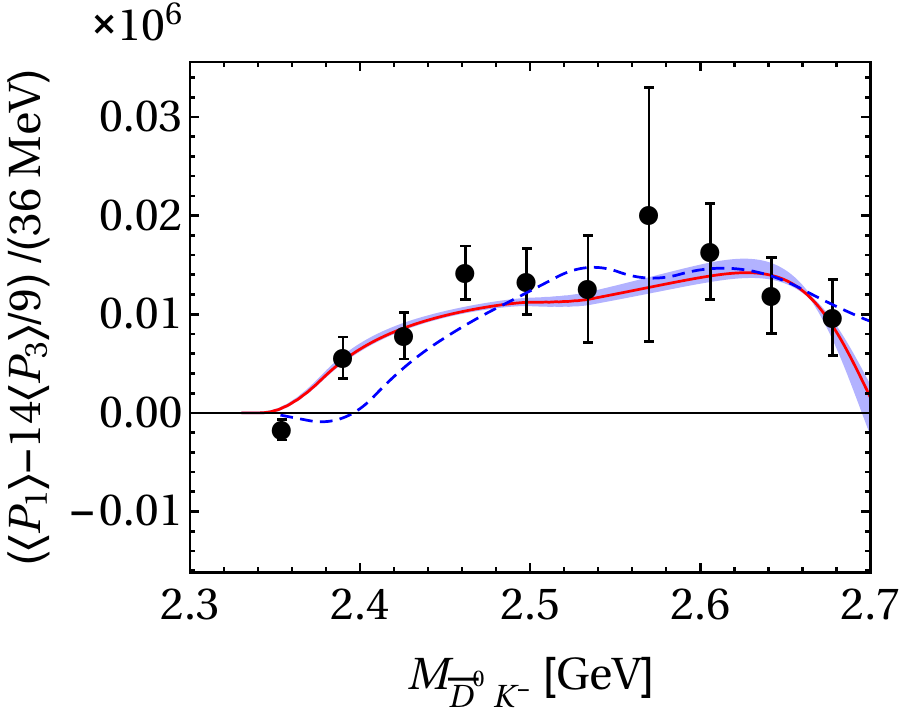} \hfill
 \includegraphics[width=0.31\linewidth]{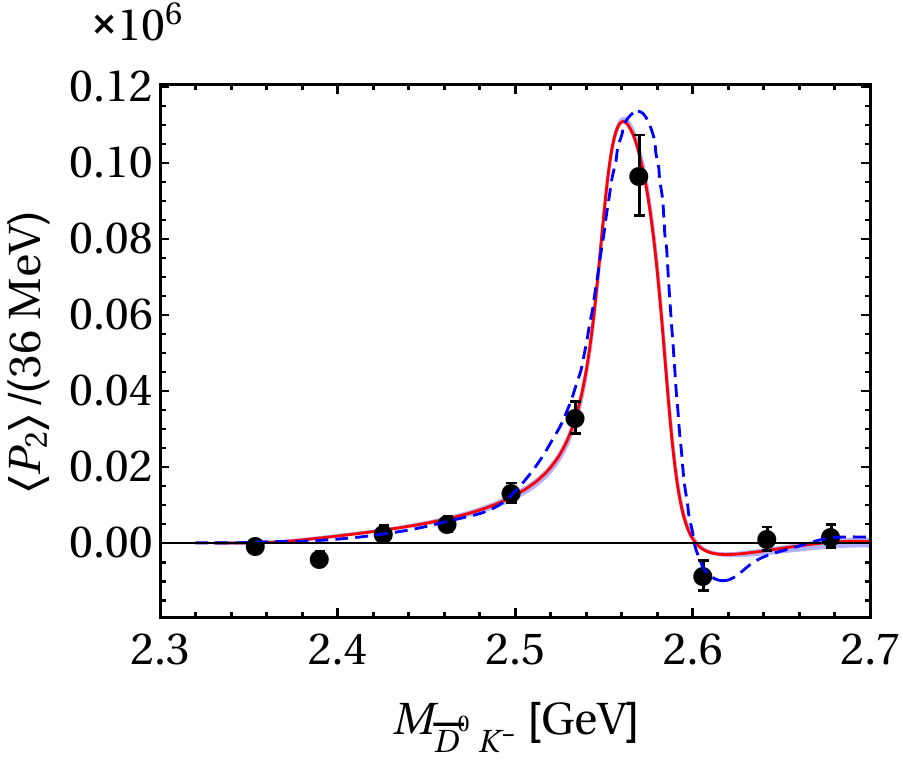}
 \end{center}
  \caption{Fit to the angular moments as a
  function of the $\bar D^0K^-$ invariant mass for the process
$ B_s^0 \to \bar{D}^0 K^- \pi^+$ provided by LHCb~\cite{Aaij:2014baa}.  The solid lines
represent the present work with the bands corresponding to the one-sigma uncertainty propagated 
from the input scattering amplitudes
and the dashed lines show the LHCb fit~\cite{Aaij:2014baa}.
  \label{fig:dk}}
\end{figure*}

{

Furthermore, the data for the angular moments for $B_s^0 \to \bar{D}^0 K^-
\pi^+$~\cite{Aaij:2014baa} can be easily reproduced in the same framework with
the $\bar D \bar K$ interaction fixed from Ref.~\cite{Liu:2012zya} again, which has
the $\bar D_{s0}^*(2317)$ as a dynamically generated state. 
We focus on the angular moments as functions of the $\bar D^0 K^-$ 
invariant mass which were measured in the LHCb experiment~\cite{Aaij:2014baa}. 
The decay mechanism is similar to the one in Fig.~\ref{fig:feyn}, and the final 
state $\bar D^0 K^-$ can be generated from $\bar D^0 K^-$, $D^- \bar K^0$, 
$\bar D_s\eta$ and $\bar D_s\pi^0$ intermediate states. Considering isospin 
symmetry, the $S$-wave part of the decay amplitude for this process can be 
written as
\begin{eqnarray}
  \A_{0}(s) \al=\al  E_K \bigg[ C + \frac12(C+A) G_1(s) T_{11}^0(s)  \nonumber\\  
  \al\al +\frac12(C-A) G_1(s) 
T_{11}^1(s) \bigg] \nonumber\\ 
\al\al- \frac1{\sqrt{3}} \left( \frac32 B-C\right) E_\eta 
G_2(s) T_{21}^0(s)\,,
\label{eq:amp2}
\end{eqnarray}
where $C=\sqrt{2}(c_2+c_4)/F$ in terms of the LECs in Eq.~\eqref{eq:Lag}, the channel labels 
1 and 2 refer to the $DK$ and $D_s\eta$ channels, respectively, and the 
superscript of the $T$-matrix refers to the isospin. Note that the Lagrangian in 
 Eq.~\eqref{eq:Lag} does not yield a term contributing to $B_s^0 \to \bar{D}_s 
\pi^0 \pi^+$. Taking the central values of $B/A$ and the subtraction constant 
$a_A$ as determined from fitting to the $B^-\to D^+\pi^-\pi^-$ data, 
there is only one free parameter in the $S$-wave amplitude, which is 
$C/A$ (we choose $A$ to serve as the normalization constant for the $S$-wave 
contribution). For the $P$- and $D$-waves, we again take the same BW 
resonances as the LHCb analysis~\cite{Aaij:2014baa}, {\it i.e.}, $D_s^*$ and 
$D_s^*(2700)$ for the $P$-wave and $D_{s2}(2573)$ for the $D$-wave with their 
masses and widths fixed to the central values in Ref.~\cite{Aaij:2014baa}. The 
best fits to the angular moments $\langle P_0\rangle$, $\langle P_2\rangle$ and 
$\langle P_1\rangle - 14 \langle P_3\rangle/9$ for the LHCb $B_s^0 \to 
\bar{D}^0 K^-
\pi^+$ data~\cite{Aaij:2014baa} up to 2.7~GeV leads to $\chi^2/{\rm d.o.f.}=1.6$, and the only free parameter in the $S$-wave amplitude is determined to be $C/A=4.8^{+3.4}_{-1.7}$. A comparison of the fit to the data is shown in Fig.~\ref{fig:dk}.
}

\medskip

In summary, we have demonstrated that amplitudes fixed from QCD inputs
for the Goldstone-Boson scattering off charm mesons not only resolve some longstanding
puzzles in charm-meson spectroscopy but also are at the same time fully consistent
with recent LHCb data on $B$ decays, which provide by far the most precise experimental
information for the $D\pi$
system.
The amplitudes have  a pole corresponding to the $D_{s0}^*(2317)$ in the
isoscalar strangeness $S=1$ channel, and two poles in the $I=1/2$ nonstrange
channel~\cite{Albaladejo:2016lbb}. The latter pair of poles should replace the lowest $J^P=0^+$ charm nonstrange meson,
 $D_0^*(2400)$, listed in the RPP~\cite{Patrignani:2016xqp}. Similarly, the broad $D_1(2430)$ listed in RPP should also be replaced by two $J^P=1^+$ states. 

It should be stressed that the observation that certain scattering amplitudes
employ poles does not necessarily imply that the corresponding states need to be
interpreted as molecular states.
However, the $S$-wave molecular admixture of a near-threshold state can be quantified
from the scattering length directly~\cite{Weinberg:1965zz}. Applying this argument to the $DK$
scattering length in the $D_{s0}^*(2317)$ channel, predicted in
Ref.~\cite{Liu:2012zya} and determined using LQCD~\cite{Mohler:2013rwa},
reveals that the molecular component of the $D_{s0}^*(2317)$
is larger than 70\%, a conclusion
confirmed later in Ref.~\cite{Torres:2014vna,Bali:2017pdv,Albaladejo:2016hae} for both 
the $D_{s0}^*(2317)$ and the $D_{s1}(2460)$.
All the poles listed in Tables~\ref{tab:masses_cs} and \ref{tab:poles} are spin-flavor partners,  due to approximate QCD symmetries. 
Therefore, they
should be envisioned as to have the same origin, i.e., hadronic molecules generated from
coupled-channel two-hadron chiral dynamics.

Treating other narrow heavy mesons, such as the
$D_1(2420)$ and the $D_2(2460)$, as matter fields leads to additional
molecular states such as the $J^P=1^-$
$D_{s1}^*(2860)$~\cite{Guo:2011dd} and its
partners. In fact, the interactions of Goldstone bosons with matter fields are relatively weak at low energies because of the chiral symmetry of QCD. Even though, hadronic molecular states can be still  generated. One would expect that the $S$-wave attractive interaction of other hadrons with heavy mesons, not suppressed by chiral symmetry, may produce hadron-hadron states as well, analogous to nuclei. These states are not the exclusive origin of higher resonances, but they are important contributors to the hadron zoo. Given more and more $S$-wave thresholds at higher energies, quark models 
are expected to become less and less reliable.

We therefore conclude that the long accepted paradigm underlying open-flavor heavy
meson spectroscopy that identifies all ground states with $c\bar q$ or $b\bar q$
quark model states, is no longer tenable. In a broader view, the hadron spectrum must be viewed
as more than a  collection of quark model states, but rather as a manifestation of a more complex dynamics
that leads to an intricate pattern of various types of states that can only be understood by a joint effort from experiment, LQCD and phenomenology.
We close the paper by summarising a few suggestions that will provide further, non-trivial tests of the scenario proposed here:
\begin{itemize}
  \item Measuring the angular moments, in particular $\langle P_1\rangle -14\langle P_3\rangle/9$, with unprecedented accuracy for the $B\to D^{(*)}\pi\pi$ and $B\to D_s^{(*)}\bar K\pi$ reactions. This can be done at LHCb and Belle-II. We expect to see nontrivial cusp structures at the $D^{(*)}\eta$ and $D_s^{(*)}\bar K$ thresholds in the former, and near-threshold enhancement in the $D_s^{(*)}\bar K$ spectrum in the latter~\cite{Albaladejo:2016lbb}.
  \item Measuring the hadronic width of the $D_{s0}^*(2317)$, predicted to be of about 100~keV in the molecular scenario~\cite{Lutz:2007sk,Liu:2012zya}, while much smaller otherwise. This will be measured by the $\overline{\text{P}}$ANDA experiment.
  \item Checking the existence of the sextet pole in LQCD with a relatively large SU(3) symmetric quark mass.
  \item Searching for the predicted analogous bottom positive-parity mesons both experimentally and in LQCD.
\end{itemize}

\medskip

\begin{acknowledgments}
We acknowledge Tim Gershon, Jonas Rademacker, Mark Peter Whitehead and Yan-Xi Zhang for discussions on the LHCb data, and thank Sin\'ead Ryan and David Wilson for communications regarding the LQCD calculation carried out in Ref.~\cite{Moir:2016srx}. 
This work is partially supported
by the National Natural Science Foundation of China (NSFC) and Deutsche Forschungsgemeinschaft (DFG) through 
funds provided to the Sino--German Collaborative Research Center ``Symmetries and the
Emergence of Structure in QCD'' (NSFC Grant No.~11621131001,
DFG Grant No.~TRR110), by the NSFC (Grant No.~11747601), by the Thousand Talents Plan for Young
Professionals, by the CAS Key Research Program of Frontier Sciences
(Grant No.~QYZDB-SSW-SYS013), by the CAS Key Research Program (Grant
No. XDPB09),  by the CAS President's
International Fellowship Initiative (PIFI) (Grant No.~2018DM0034), by the CAS Center for Excellence in Particle Physics (CCEPP), by
Spanish Ministerio de Economía y Competitividad
and the European Regional Development Fund
under contracts FIS2014-51948-C2-1-P, FPA2016-77313-P, FIS2017-84038-C2-1-P and SEV-
2014-0398, by Generalitat Valenciana under contract
PROMETEOII/2014/0068, and by the “Ayudas para contratos predoctorales para la formación de doctores” program (BES-2015-072049) from the Spanish MINECO and ESF.
\end{acknowledgments}

\begin{appendix}

{

\section{Effective Lagrangian}
\label{app:Lagrangian}
\renewcommand{\theequation}{\thesection.\arabic{equation}}
\setcounter{equation}{0}

Here, we discuss briefly the effective Lagrangian for the weak decays
$\bar B$ to $D$ with the emission of two light pseudoscalar mesons, induced by
the Cabibbo-allowed  transition $b\to c \bar u d$.
In the phase space region near the $D\pi$ threshold, chiral symmetry puts
constraints on one of the two pions while the other one moves fast and can be
treated as a matter field. Moreover, its interaction with the other particles in
the final state can be safely neglected. Then the relevant chiral 
effective Lagrangian leading to Eq.~\eqref{eq:amp} reads,
\begin{eqnarray}
\lag_\text{eff} \al = \al \bar B \big[ c_1 \left(u_\mu t M + Mtu_\mu \right)
+ c_2 \left(u_\mu M + Mu_\mu \right)t \nonumber\\
\al\al + c_3\, t \left(u_\mu M+Mu_\mu \right) + c_4  \left(u_\mu \langle M t\rangle + M \langle u_\mu t\rangle \right)
\nonumber\\
\al\al + c_5\, t \langle M u_\mu\rangle + c_6 \langle \left(Mu_\mu+u_\mu
M\right)t\rangle
\big] \partial^\mu D^\dag \, .
\label{eq:Lag}
\end{eqnarray}
Here, $\bar B= (B^-, \bar B^0, \bar B_s^0)$ and $D= (D^0, D^+, D_s^+)$ are the
fields for bottom and charm mesons, $\langle\ldots\rangle$ denotes the trace in 
the SU(3) light-flavor space, and $u_\mu = i ( u^\dag \partial_\mu u -
u\partial_\mu u^\dag )$ is the axial current derived from chiral symmetry. 
The Goldstone Bosons are represented non-linearily via $u=\exp\left(i\phi/(\sqrt{2}F)\right)$, 
with
\begin{equation}
 \phi = \begin{pmatrix}
\frac{1}{\sqrt{2}}\pi^0 +\frac{1}{\sqrt{6}}\eta & \pi^+ & K^+ \\
\pi^- & -\frac{1}{\sqrt{2}}\pi^0 +\frac{1}{\sqrt{6}}\eta & K^0 \\
K^- & \bar{K}^0 & -\frac{2}{\sqrt{6}}\eta
\label{eq:phi}
\end{pmatrix} \ ,
\end{equation}
in terms of the pions ($\pi^\pm, \pi^0$), the kaons ($K^\pm, K^0, \bar{K}^0$) and the
$\eta$, and $F$ denotes the Goldstone-Boson decay constant in the chiral limit.
In addition, $t=uHu^\dag$ is a spurion field with~\cite{Savage:1989ub}
\begin{equation}
  H = \begin{pmatrix}
0 & 0 & 0 \\
1 & 0 & 0 \\
0 & 0 & 0
\end{pmatrix},
\end{equation}
for Cabibbo-allowed decays.
The matter field $M$, having the same form as $\phi$, describes the
fast moving light meson. The $c_i$ $(i=1,\ldots,6$) are low-energy constants (LECs).
This effective Lagrangian considers both chiral, for the regime with soft Goldstone Bosons, and SU(3) constraints, the latter of which has been considered in Ref.~\cite{Savage:1989ub}. In terms of the LECs in the above Lagrangian, the parameters $A$ and $B$ in Eq.~\eqref{eq:amp} can be expressed as $A=\sqrt{2}(c_1+c_4)/F$ and $B=2\sqrt{2}(c_2+c_6)/(3F)$.
}

\end{appendix}


\begin{thebibliography}{99} 

%\cite{GellMann:1964nj}
\bibitem{GellMann:1964nj} 
  M.~Gell-Mann,
  %``A Schematic Model of Baryons and Mesons,''
  Phys.\ Lett.\  {\bf 8}, 214 (1964).
%   doi:10.1016/S0031-9163(64)92001-3
  %%CITATION = doi:10.1016/S0031-9163(64)92001-3;%%
  %2633 citations counted in INSPIRE as of 20 Dec 2017


%\cite{Godfrey:1985xj}
\bibitem{Godfrey:1985xj} 
  S.~Godfrey and N.~Isgur,
  %``Mesons in a Relativized Quark Model with Chromodynamics,''
  Phys.\ Rev.\ D {\bf 32}, 189 (1985).
%   doi:10.1103/PhysRevD.32.189
  %%CITATION = doi:10.1103/PhysRevD.32.189;%%
  %2394 citations counted in INSPIRE as of 20 Dec 2017


%\cite{Aubert:2003fg}
\bibitem{Aubert:2003fg} 
  B.~Aubert {\it et al.} [BaBar Collaboration],
  %``Observation of a narrow meson decaying to $D_s^+ \pi^0$ at a mass of 2.32-GeV/c$^2$,''
  Phys.\ Rev.\ Lett.\  {\bf 90}, 242001 (2003)
%   doi:10.1103/PhysRevLett.90.242001
  [hep-ex/0304021].
  %%CITATION = doi:10.1103/PhysRevLett.90.242001;%%
  %804 citations counted in INSPIRE as of 20 Dec 2017


%\cite{Besson:2003cp}
\bibitem{Besson:2003cp} 
  D.~Besson {\it et al.} [CLEO Collaboration],
  %``Observation of a narrow resonance of mass 2.46-GeV/c**2 decaying to D*+(s) pi0 and confirmation of the D*(sJ)(2317) state,''
  Phys.\ Rev.\ D {\bf 68}, 032002 (2003)
  Erratum: [Phys.\ Rev.\ D {\bf 75}, 119908 (2007)]
%   doi:10.1103/PhysRevD.68.032002, 10.1103/PhysRevD.75.119908
  [hep-ex/0305100].
  %%CITATION = doi:10.1103/PhysRevD.68.032002, 10.1103/PhysRevD.75.119908;%%
  %566 citations counted in INSPIRE as of 20 Dec 2017


%\cite{Chen:2016qju}
\bibitem{Chen:2016qju} 
  H.~X.~Chen, W.~Chen, X.~Liu and S.~L.~Zhu,
  %``The hidden-charm pentaquark and tetraquark states,''
  Phys.\ Rept.\  {\bf 639}, 1 (2016)
%   doi:10.1016/j.physrep.2016.05.004
  [arXiv:1601.02092 [hep-ph]].
  %%CITATION = doi:10.1016/j.physrep.2016.05.004;%%
  %220 citations counted in INSPIRE as of 20 Dec 2017


%\cite{Chen:2016spr}
\bibitem{Chen:2016spr} 
  H.~X.~Chen, W.~Chen, X.~Liu, Y.~R.~Liu and S.~L.~Zhu,
  %``A review of the open charm and open bottom systems,''
  Rept.\ Prog.\ Phys.\  {\bf 80}, 076201 (2017)
%   doi:10.1088/1361-6633/aa6420
  [arXiv:1609.08928 [hep-ph]].
  %%CITATION = doi:10.1088/1361-6633/aa6420;%%
  %48 citations counted in INSPIRE as of 20 Dec 2017


%\cite{Lebed:2016hpi}
\bibitem{Lebed:2016hpi} 
  R.~F.~Lebed, R.~E.~Mitchell and E.~S.~Swanson,
  %``Heavy-Quark QCD Exotica,''
  Prog.\ Part.\ Nucl.\ Phys.\  {\bf 93}, 143 (2017)
%   doi:10.1016/j.ppnp.2016.11.003
  [arXiv:1610.04528 [hep-ph]].
  %%CITATION = doi:10.1016/j.ppnp.2016.11.003;%%
  %49 citations counted in INSPIRE as of 20 Dec 2017


%\cite{Esposito:2016noz}
\bibitem{Esposito:2016noz} 
  A.~Esposito, A.~Pilloni and A.~D.~Polosa,
  %``Multiquark Resonances,''
  Phys.\ Rept.\  {\bf 668}, 1 (2016)
%   doi:10.1016/j.physrep.2016.11.002
  [arXiv:1611.07920 [hep-ph]].
  %%CITATION = doi:10.1016/j.physrep.2016.11.002;%%
  %58 citations counted in INSPIRE as of 20 Dec 2017


%\cite{Guo:2017jvc}
\bibitem{Guo:2017jvc} 
  F.-K.~Guo, C.~Hanhart, U.-G.~Mei{\ss}ner, Q.~Wang, Q.~Zhao and B.~S.~Zou,
  %``Hadronic molecules,''
  Rev. Mod. Phys. {\bf 90}, 015004 (2018)
  [arXiv:1705.00141 [hep-ph]].
  %%CITATION = ARXIV:1705.00141;%%
  %57 citations counted in INSPIRE as of 20 Dec 2017


%\cite{Ali:2017jda}
\bibitem{Ali:2017jda} 
  A.~Ali, J.~S.~Lange and S.~Stone,
  %``Exotics: Heavy Pentaquarks and Tetraquarks,''
  Prog.\ Part.\ Nucl.\ Phys.\  {\bf 97}, 123 (2017)
%   doi:10.1016/j.ppnp.2017.08.003
  [arXiv:1706.00610 [hep-ph]].
  %%CITATION = doi:10.1016/j.ppnp.2017.08.003;%%
  %24 citations counted in INSPIRE as of 20 Dec 2017


%\cite{Olsen:2017bmm}
\bibitem{Olsen:2017bmm} 
  S.~L.~Olsen, T.~Skwarnicki and D.~Zieminska,
  %``NonStandard Heavy Mesons and Baryons: Experimental Evidence,''
  Rev. Mod. Phys. {\bf 90}, 015003 (2018)
  [arXiv:1708.04012 [hep-ph]].
  %%CITATION = ARXIV:1708.04012;%%
  %16 citations counted in INSPIRE as of 20 Dec 2017


%\cite{Cahn:2003cw}
\bibitem{Cahn:2003cw} 
  R.~N.~Cahn and J.~D.~Jackson,
  %``Spin orbit and tensor forces in heavy quark light quark mesons: Implications of the new D(s) state at 2.32-GeV,''
  Phys.\ Rev.\ D {\bf 68}, 037502 (2003)
%   doi:10.1103/PhysRevD.68.037502
  [hep-ph/0305012].
  %%CITATION = doi:10.1103/PhysRevD.68.037502;%%
  %147 citations counted in INSPIRE as of 20 Dec 2017


%\cite{Barnes:2003dj}
\bibitem{Barnes:2003dj} 
  T.~Barnes, F.~E.~Close and H.~J.~Lipkin,
  %``Implications of a DK molecule at 2.32-GeV,''
  Phys.\ Rev.\ D {\bf 68}, 054006 (2003)
%   doi:10.1103/PhysRevD.68.054006
  [hep-ph/0305025].
  %%CITATION = doi:10.1103/PhysRevD.68.054006;%%
  %337 citations counted in INSPIRE as of 20 Dec 2017


%\cite{vanBeveren:2003kd}
\bibitem{vanBeveren:2003kd} 
  E.~van Beveren and G.~Rupp,
  %``Observed D(s)(2317) and tentative D(2030) as the charmed cousins of the light scalar nonet,''
  Phys.\ Rev.\ Lett.\  {\bf 91}, 012003 (2003)
%   doi:10.1103/PhysRevLett.91.012003
  [hep-ph/0305035].
  %%CITATION = doi:10.1103/PhysRevLett.91.012003;%%
  %297 citations counted in INSPIRE as of 20 Dec 2017


%\cite{Chen:2004dy}
\bibitem{Chen:2004dy} 
  Y.-Q.~Chen and X.-Q.~Li,
  %``A Comprehensive four-quark interpretation of D(s)(2317), D(s)(2457) and D(s)(2632),''
  Phys.\ Rev.\ Lett.\  {\bf 93}, 232001 (2004)
%   doi:10.1103/PhysRevLett.93.232001
  [hep-ph/0407062].
  %%CITATION = doi:10.1103/PhysRevLett.93.232001;%%
  %83 citations counted in INSPIRE as of 20 Dec 2017


%\cite{Bardeen:2003kt}
\bibitem{Bardeen:2003kt} 
  W.~A.~Bardeen, E.~J.~Eichten and C.~T.~Hill,
  %``Chiral multiplets of heavy - light mesons,''
  Phys.\ Rev.\ D {\bf 68}, 054024 (2003)
%   doi:10.1103/PhysRevD.68.054024
  [hep-ph/0305049].
  %%CITATION = doi:10.1103/PhysRevD.68.054024;%%
  %429 citations counted in INSPIRE as of 20 Dec 2017


%\cite{Nowak:2003ra}
\bibitem{Nowak:2003ra} 
  M.~A.~Nowak, M.~Rho and I.~Zahed,
  %``Chiral doubling of heavy light hadrons: BABAR 2317-MeV/c**2 and CLEO 2463-MeV/c**2 discoveries,''
  Acta Phys.\ Polon.\ B {\bf 35}, 2377 (2004)
  [hep-ph/0307102].
  %%CITATION = HEP-PH/0307102;%%
  %145 citations counted in INSPIRE as of 20 Dec 2017


%\cite{Link:2003bd}
\bibitem{Link:2003bd} 
  J.~M.~Link {\it et al.} [FOCUS Collaboration],
  %``Measurement of masses and widths of excited charm mesons D(2)* and evidence for broad states,''
  Phys.\ Lett.\ B {\bf 586}, 11 (2004)
%   doi:10.1016/j.physletb.2004.02.017
  [hep-ex/0312060].
  %%CITATION = doi:10.1016/j.physletb.2004.02.017;%%
  %132 citations counted in INSPIRE as of 20 Dec 2017


%\cite{Abe:2003zm}
\bibitem{Abe:2003zm} 
  K.~Abe {\it et al.} [Belle Collaboration],
  %``Study of B- ---> D**0 pi- (D**0 ---> D(*)+ pi-) decays,''
  Phys.\ Rev.\ D {\bf 69}, 112002 (2004)
%   doi:10.1103/PhysRevD.69.112002
  [hep-ex/0307021].
  %%CITATION = doi:10.1103/PhysRevD.69.112002;%%
  %332 citations counted in INSPIRE as of 20 Dec 2017


%\cite{Mehen:2005hc}
\bibitem{Mehen:2005hc} 
  T.~Mehen and R.~P.~Springer,
  %``Even- and odd-parity charmed meson masses in heavy hadron chiral perturbation theory,''
  Phys.\ Rev.\ D {\bf 72}, 034006 (2005)
%   doi:10.1103/PhysRevD.72.034006
  [hep-ph/0503134].
  %%CITATION = doi:10.1103/PhysRevD.72.034006;%%
  %35 citations counted in INSPIRE as of 20 Dec 2017


%\cite{Colangelo:2012xi}
\bibitem{Colangelo:2012xi}
  P.~Colangelo, F.~De Fazio, F.~Giannuzzi and S.~Nicotri,
  %``New meson spectroscopy with open charm and beauty,''
  Phys.\ Rev.\ D {\bf 86}, 054024 (2012)
%   doi:10.1103/PhysRevD.86.054024
  [arXiv:1207.6940 [hep-ph]].
  %%CITATION = doi:10.1103/PhysRevD.86.054024;%%
  %86 citations counted in INSPIRE as of 06 Dec 2017

\bibitem{Alhakami:2016zqx} 
  M.~H.~Alhakami,
  %``Mass Spectra of Heavy-Light Mesons in Heavy Hadron Chiral Perturbation Theory,''
  Phys.\ Rev.\ D {\bf 93}, 094007 (2016)
%   doi:10.1103/PhysRevD.93.094007
  [arXiv:1603.08848 [hep-ph]].

%\cite{Cheng:2017oqh}
\bibitem{Cheng:2017oqh}
  H.-Y.~Cheng and F.-S.~Yu,
  %``Masses of Scalar and Axial-Vector B Mesons Revisited,''
  Eur.\ Phys.\ J.\ C {\bf 77}, 668 (2017)
%   doi:10.1140/epjc/s10052-017-5252-4
  [arXiv:1704.01208 [hep-ph]].
  %%CITATION = doi:10.1140/epjc/s10052-017-5252-4;%%

%\cite{Patrignani:2016xqp}
\bibitem{Patrignani:2016xqp} 
%   C.~Patrignani {\it et al.} [Particle Data Group],
%   %``Review of Particle Physics,''
%   Chin.\ Phys.\ C {\bf 40}, 100001 (2016).
   M.~Tanabashi {\it et al.} [Particle Data Group],
  %``Review of Particle Physics,''
  Phys.\ Rev. D {\bf 98}, 030001 (2018).



%\cite{Kolomeitsev:2003ac}
\bibitem{Kolomeitsev:2003ac} 
  E.~E.~Kolomeitsev and M.~F.~M.~Lutz,
  %``On Heavy light meson resonances and chiral symmetry,''
  Phys.\ Lett.\ B {\bf 582}, 39 (2004)
%   doi:10.1016/j.physletb.2003.10.118
  [hep-ph/0307133].
  %%CITATION = doi:10.1016/j.physletb.2003.10.118;%%
  %192 citations counted in INSPIRE as of 20 Dec 2017


%\cite{Guo:2006fu}
\bibitem{Guo:2006fu} 
  F.-K.~Guo, P.-N.~Shen, H.-C.~Chiang, R.-G.~Ping and B.-S.~Zou,
  %``Dynamically generated 0+ heavy mesons in a heavy chiral unitary approach,''
  Phys.\ Lett.\ B {\bf 641}, 278 (2006)
%   doi:10.1016/j.physletb.2006.08.064
  [hep-ph/0603072].
  %%CITATION = doi:10.1016/j.physletb.2006.08.064;%%
  %149 citations counted in INSPIRE as of 20 Dec 2017


%\cite{Hofmann:2003je}
\bibitem{Hofmann:2003je} 
  J.~Hofmann and M.~F.~M.~Lutz,
  %``Open charm meson resonances with negative strangeness,''
  Nucl.\ Phys.\ A {\bf 733}, 142 (2004)
%   doi:10.1016/j.nuclphysa.2003.12.013
  [hep-ph/0308263].
  %%CITATION = doi:10.1016/j.nuclphysa.2003.12.013;%%
  %116 citations counted in INSPIRE as of 20 Dec 2017


%\cite{Guo:2006rp}
\bibitem{Guo:2006rp} 
  F.-K.~Guo, P.-N.~Shen and H.-C.~Chiang,
  %``Dynamically generated 1+ heavy mesons,''
  Phys.\ Lett.\ B {\bf 647}, 133 (2007)
%   doi:10.1016/j.physletb.2007.01.050
  [hep-ph/0610008].
  %%CITATION = doi:10.1016/j.physletb.2007.01.050;%%
  %81 citations counted in INSPIRE as of 20 Dec 2017


%\cite{Gamermann:2006nm}
\bibitem{Gamermann:2006nm} 
  D.~Gamermann, E.~Oset, D.~Strottman and M.~J.~Vicente Vacas,
  %``Dynamically generated open and hidden charm meson systems,''
  Phys.\ Rev.\ D {\bf 76}, 074016 (2007)
%   doi:10.1103/PhysRevD.76.074016
  [hep-ph/0612179].
  %%CITATION = doi:10.1103/PhysRevD.76.074016;%%
  %187 citations counted in INSPIRE as of 20 Dec 2017


%\cite{Flynn:2007ki}
\bibitem{Flynn:2007ki} 
  J.~M.~Flynn and J.~Nieves,
  %``Elastic s-wave B pi, D pi, D K and K pi scattering from lattice calculations of scalar form-factors in semileptonic decays,''
  Phys.\ Rev.\ D {\bf 75}, 074024 (2007)
%   doi:10.1103/PhysRevD.75.074024
  [hep-ph/0703047].
  %%CITATION = doi:10.1103/PhysRevD.75.074024;%%
  %32 citations counted in INSPIRE as of 20 Dec 2017


%\cite{Guo:2009ct}
\bibitem{Guo:2009ct} 
  F.-K.~Guo, C.~Hanhart and U.-G.~Mei{\ss}ner,
  %``Interactions between heavy mesons and Goldstone bosons from chiral dynamics,''
  Eur.\ Phys.\ J.\ A {\bf 40}, 171 (2009)
%   doi:10.1140/epja/i2009-10762-1
  [arXiv:0901.1597 [hep-ph]].
  %%CITATION = doi:10.1140/epja/i2009-10762-1;%%
  %82 citations counted in INSPIRE as of 20 Dec 2017


%\cite{Liu:2012zya}
\bibitem{Liu:2012zya} 
  L.~Liu, K.~Orginos, F.-K.~Guo, C.~Hanhart and U.-G.~Mei{\ss}ner,
  %``Interactions of charmed mesons with light pseudoscalar mesons from lattice QCD and implications on the nature of the $D_{s0}^*(2317)$,''
  Phys.\ Rev.\ D {\bf 87}, 014508 (2013)
%   doi:10.1103/PhysRevD.87.014508
  [arXiv:1208.4535 [hep-lat]].
  %%CITATION = doi:10.1103/PhysRevD.87.014508;%%
  %66 citations counted in INSPIRE as of 20 Dec 2017


%\cite{Altenbuchinger:2013vwa}
\bibitem{Altenbuchinger:2013vwa} 
  M.~Altenbuchinger, L.-S.~Geng and W.~Weise,
  %``Scattering lengths of Nambu-Goldstone bosons off $D$ mesons and dynamically generated heavy-light mesons,''
  Phys.\ Rev.\ D {\bf 89}, 014026 (2014)
%   doi:10.1103/PhysRevD.89.014026
  [arXiv:1309.4743 [hep-ph]].
  %%CITATION = doi:10.1103/PhysRevD.89.014026;%%
  %34 citations counted in INSPIRE as of 20 Dec 2017

%\cite{Du:2017ttu}
\bibitem{Du:2017ttu} 
  M.-L.~Du, F.-K.~Guo, U.-G.~Mei{\ss}ner and D.-L.~Yao,
  %``Study of open-charm $0^+$ states in unitarized chiral effective theory with one-loop potentials,''
  Eur.\ Phys.\ J.\ C {\bf 77}, 728 (2017)
%   doi:10.1140/epjc/s10052-017-5287-6
  [arXiv:1703.10836 [hep-ph]].
  %%CITATION = doi:10.1140/epjc/s10052-017-5287-6;%%
  %3 citations counted in INSPIRE as of 20 Dec 2017


%\cite{Truong:1988zp}
\bibitem{Truong:1988zp} 
  T.~N.~Truong,
  %``Chiral Perturbation Theory and Final State Theorem,''
  Phys.\ Rev.\ Lett.\  {\bf 61}, 2526 (1988).
%   doi:10.1103/PhysRevLett.61.2526
  %%CITATION = doi:10.1103/PhysRevLett.61.2526;%%
  %260 citations counted in INSPIRE as of 20 Dec 2017

\bibitem{Pelaez:2015qba} 
  J.~R.~Pel\'aez,
  %``From controversy to precision on the sigma meson: a review on the status of the non-ordinary $f_0(500)$ resonance,''
  Phys.\ Rept.\  {\bf 658}, 1 (2016)
%   doi:10.1016/j.physrep.2016.09.001
  [arXiv:1510.00653 [hep-ph]].
  
  %\cite{Gasser:1990bv}
\bibitem{Gasser:1990bv}
  J.~Gasser and U.-G.~Mei{\ss}ner,
  %``Chiral expansion of pion form-factors beyond one loop,''
  Nucl.\ Phys.\ B {\bf 357} (1991) 90.

\bibitem{Bruns:2017gix}
  P.~C.~Bruns and M.~Mai,
  %``Chiral symmetry constraints on resonant amplitudes,''
  Phys.\ Lett.\ B {\bf 778} (2018) 43
%  doi:10.1016/j.physletb.2018.01.006
  [arXiv:1707.08983 [hep-lat]].

%\cite{Albaladejo:2016lbb}
\bibitem{Albaladejo:2016lbb} 
  M.~Albaladejo, P.~Fernandez-Soler, F.-K.~Guo and J.~Nieves,
  %``Two-pole structure of the $D^\ast_0(2400)$,''
  Phys.\ Lett.\ B {\bf 767}, 465 (2017)
%   doi:10.1016/j.physletb.2017.02.036
  [arXiv:1610.06727 [hep-ph]].
  %%CITATION = doi:10.1016/j.physletb.2017.02.036;%%
  %8 citations counted in INSPIRE as of 20 Dec 2017


%\cite{Moir:2016srx}
\bibitem{Moir:2016srx} 
  G.~Moir, M.~Peardon, S.~M.~Ryan, C.~E.~Thomas and D.~J.~Wilson,
  %``Coupled-Channel $D\pi$, $D\eta$ and $D_{s}\bar{K}$ Scattering from Lattice QCD,''
  JHEP {\bf 1610}, 011 (2016)
%   doi:10.1007/JHEP10(2016)011
  [arXiv:1607.07093 [hep-lat]].
  %%CITATION = doi:10.1007/JHEP10(2016)011;%%
  %31 citations counted in INSPIRE as of 20 Dec 2017


%\cite{Neubert:1993mb}
\bibitem{Neubert:1993mb} 
  M.~Neubert,
  %``Heavy quark symmetry,''
  Phys.\ Rept.\  {\bf 245}, 259 (1994)
%   doi:10.1016/0370-1573(94)90091-4
  [hep-ph/9306320].
  %%CITATION = doi:10.1016/0370-1573(94)90091-4;%%
  %1414 citations counted in INSPIRE as of 20 Dec 2017

%\cite{Lang:2015hza}
\bibitem{Lang:2015hza} 
  C.~B.~Lang, D.~Mohler, S.~Prelovsek and R.~M.~Woloshyn,
  %``Predicting positive parity B$_s$ mesons from lattice QCD,''
  Phys.\ Lett.\ B {\bf 750}, 17 (2015)
%   doi:10.1016/j.physletb.2015.08.038
  [arXiv:1501.01646 [hep-lat]].
  %%CITATION = doi:10.1016/j.physletb.2015.08.038;%%
  %36 citations counted in INSPIRE as of 20 Dec 2017
  
\bibitem{Davidprivate}
David Wilson, private communication.


\bibitem{Hanhart:2014ssa}
  C.~Hanhart, J.~R.~Pel\'aez and G.~R\'ios,
  %``Remarks on pole trajectories for resonances,''
  Phys.\ Lett.\ B {\bf 739} (2014) 375
%   doi:10.1016/j.physletb.2014.11.011
  [arXiv:1407.7452 [hep-ph]].
  
  \bibitem{Hanhart:2008mx} 
  C.~Hanhart, J.~R.~Pel\'aez and G.~R\'ios,
  %``Quark mass dependence of the rho and sigma from dispersion relations and Chiral Perturbation Theory,''
  Phys.\ Rev.\ Lett.\  {\bf 100}, 152001 (2008)
%   doi:10.1103/PhysRevLett.100.152001
  [arXiv:0801.2871 [hep-ph]].
  
  \bibitem{Albaladejo:2012te} 
  M.~Albaladejo and J.~A.~Oller,
  %``On the size of the sigma meson and its nature,''
  Phys.\ Rev.\ D {\bf 86}, 034003 (2012)
%   doi:10.1103/PhysRevD.86.034003
  [arXiv:1205.6606 [hep-ph]].

%\cite{Bali:2017pdv}
\bibitem{Bali:2017pdv} 
  G.~S.~Bali, S.~Collins, A.~Cox and A.~Sch{\"a}fer,
  %``Masses and decay constants of the $D_{s0}^*(2317)$ and $D_{s1}(2460)$ from $N_f=2$ lattice QCD close to the physical point,''
  Phys.\ Rev.\ D {\bf 96}, 074501 (2017)
%   doi:10.1103/PhysRevD.96.074501
  [arXiv:1706.01247 [hep-lat]].
  %%CITATION = doi:10.1103/PhysRevD.96.074501;%%
  %2 citations counted in INSPIRE as of 20 Dec 2017

%\cite{Aaij:2016fma}
\bibitem{Aaij:2016fma} 
  R.~Aaij {\it et al.} [LHCb Collaboration],
  %``Amplitude analysis of $B^{-} \to D^{+} \pi^{-} \pi^{-}$ decays,''
  Phys.\ Rev.\ D {\bf 94}, 072001 (2016)
%   doi:10.1103/PhysRevD.94.072001
  [arXiv:1608.01289 [hep-ex]].
  %%CITATION = doi:10.1103/PhysRevD.94.072001;%%
  %15 citations counted in INSPIRE as of 20 Dec 2017



%\cite{Savage:1989ub}
\bibitem{Savage:1989ub} 
  M.~J.~Savage and M.~B.~Wise,
  %``SU(3) Predictions for Nonleptonic B Meson Decays,''
  Phys.\ Rev.\ D {\bf 39}, 3346 (1989)
  Erratum: [Phys.\ Rev.\ D {\bf 40}, 3127 (1989)].
%   doi:10.1103/PhysRevD.39.3346, 10.1103/PhysRevD.40.3127
  %%CITATION = doi:10.1103/PhysRevD.39.3346, 10.1103/PhysRevD.40.3127;%%
  %146 citations counted in INSPIRE as of 20 Dec 2017


%\cite{Oller:1998zr}
\bibitem{Oller:1998zr} 
  J.~A.~Oller and E.~Oset,
  %``N/D description of two meson amplitudes and chiral symmetry,''
  Phys.\ Rev.\ D {\bf 60}, 074023 (1999)
%   doi:10.1103/PhysRevD.60.074023
  [hep-ph/9809337].
  %%CITATION = doi:10.1103/PhysRevD.60.074023;%%
  %505 citations counted in INSPIRE as of 20 Dec 2017


%\cite{Aaij:2014baa}
\bibitem{Aaij:2014baa} 
  R.~Aaij {\it et al.} [LHCb Collaboration],
  %``Dalitz plot analysis of $B_s^0 \rightarrow \bar{D}^0 K^- \pi^+$ decays,''
  Phys.\ Rev.\ D {\bf 90}, 072003 (2014)
%   doi:10.1103/PhysRevD.90.072003
  [arXiv:1407.7712 [hep-ex]].
  %%CITATION = doi:10.1103/PhysRevD.90.072003;%%
  %80 citations counted in INSPIRE as of 20 Dec 2017


%\cite{Weinberg:1965zz}
\bibitem{Weinberg:1965zz} 
  S.~Weinberg,
  %``Evidence That the Deuteron Is Not an Elementary Particle,''
  Phys.\ Rev.\  {\bf 137}, B672 (1965).
%   doi:10.1103/PhysRev.137.B672
  %%CITATION = doi:10.1103/PhysRev.137.B672;%%
  %208 citations counted in INSPIRE as of 20 Dec 2017


%\cite{Mohler:2013rwa}
\bibitem{Mohler:2013rwa} 
  D.~Mohler, C.~B.~Lang, L.~Leskovec, S.~Prelovsek and R.~M.~Woloshyn,
  %``$D_{s0}^*(2317)$ Meson and $D$-Meson-Kaon Scattering from Lattice QCD,''
  Phys.\ Rev.\ Lett.\  {\bf 111},  222001 (2013)
%   doi:10.1103/PhysRevLett.111.222001
  [arXiv:1308.3175 [hep-lat]].
  %%CITATION = doi:10.1103/PhysRevLett.111.222001;%%
  %98 citations counted in INSPIRE as of 20 Dec 2017

%\cite{Torres:2014vna}
\bibitem{Torres:2014vna}
  A.~Mart{\'i}nez Torres, E.~Oset, S.~Prelovsek and A.~Ramos,
  %``Reanalysis of lattice QCD spectra leading to the $D_{s0}^*(2317)$ and
% $D_{s1}^*(2460)$,''
  JHEP {\bf 1505}, 153 (2015)
%   doi:10.1007/JHEP05(2015)153
  [arXiv:1412.1706 [hep-lat]].
  %%CITATION = doi:10.1007/JHEP05(2015)153;%%
  %39 citations counted in INSPIRE as of 06 Dec 2017

\bibitem{Albaladejo:2016hae} 
  M.~Albaladejo, D.~Jido, J.~Nieves and E.~Oset,
  %``$D^*_{s0}(2317)$ and $\textit{DK}$ scattering in B decays from BaBar and LHCb data,''
  Eur.\ Phys.\ J.\ C {\bf 76}, 300 (2016)
%   doi:10.1140/epjc/s10052-016-4144-3
  [arXiv:1604.01193 [hep-ph]].


%\cite{Guo:2011dd}
\bibitem{Guo:2011dd} 
  F.-K.~Guo and U.-G.~Mei{\ss}ner,
  %``More kaonic bound states and a comprehensive interpretation of the $D_{sJ}$ states,''
  Phys.\ Rev.\ D {\bf 84}, 014013 (2011)
%   doi:10.1103/PhysRevD.84.014013
  [arXiv:1102.3536 [hep-ph]].
  %%CITATION = doi:10.1103/PhysRevD.84.014013;%%
  %31 citations counted in INSPIRE as of 20 Dec 2017
 
%\cite{Lutz:2007sk}
\bibitem{Lutz:2007sk} 
  M.~F.~M.~Lutz and M.~Soyeur,
  %``Radiative and isospin-violating decays of D(s)-mesons in the hadrogenesis conjecture,''
  Nucl.\ Phys.\ A {\bf 813}, 14 (2008)
%   doi:10.1016/j.nuclphysa.2008.09.003
  [arXiv:0710.1545 [hep-ph]].
  %%CITATION = doi:10.1016/j.nuclphysa.2008.09.003;%%
  %67 citations counted in INSPIRE as of 20 Dec 2017


\end{thebibliography}
\end{document}